\documentclass[twocolumn]{aastex7}

\graphicspath{{./}{figures/}}
\usepackage{amsmath}
\usepackage{multirow}    
\usepackage{array}       
\usepackage{comment}
\usepackage{natbib}
\usepackage{graphicx}
\usepackage{epstopdf}
\usepackage{placeins}
\usepackage{float}
\usepackage{siunitx}
\usepackage{longtable}
\DeclareUnicodeCharacter{FF1B}{;}
\begin{document}
\title{Constraining the Pulsar 3D Velocity Distribution: The Impact of Spin-Velocity Alignment
}

\author[0000-0002-0786-7307]{Zheng Li}
\affiliation{Institute for Frontier in Astronomy and Astrophysics \& Faculty of Arts and Sciences, Beijing Normal University, Zhuhai 519087, China}
\affiliation{School of Physics and Astronomy, Beijing Normal University, Beijing 100875, China}
\email{l.z@bnu.edu.cn}

\author[0000-0002-2187-4087]{Xiaojin Liu}
\affiliation{Institute for Frontier in Astronomy and Astrophysics \& Faculty of Arts and Sciences, Beijing Normal University, Zhuhai 519087, China}
\email{xliu@bnu.edu.cn}

\author[0000-0002-3309-415X]{Zhi-Qiang You}
\affiliation{Institute for Gravitational Wave Astronomy, Henan Academy of Sciences, Zhengzhou 450046, Henan, China}
\email{you_zhiqiang@whu.edu.cn}

\author[0000-0002-4997-045X]{Jumei Yao}
\affiliation{Xinjiang Astronomical Observatory, Chinese Academy of Sciences, Urumqi, 830011,China}
\affiliation{Key Laboratory of Radio Astronomy, Chinese Academy of Sciences, Urumqi, 830011,China}
\affiliation{Xinjiang Key Laboratory of Radio Astrophysics, Urumqi, 830011, China}
\email{yaojumei@xao.ac.cn}

\author[0000-0001-7049-6468]{Xing-Jiang Zhu}
\affiliation{Institute for Frontier in Astronomy and Astrophysics \& Faculty of Arts and Sciences, Beijing Normal University, Zhuhai 519087, China}
\email[show]{zhuxj@bnu.edu.cn}

\begin{abstract}

Quantifying the natal kick distribution of pulsars is essential for understanding supernova physics and binary evolution, yet measurements are historically limited by the lack of radial velocity data. Most previous studies rely on transverse velocities under the assumption of spatial isotropy. In this work, we reconstruct the intrinsic three-dimensional (3D) velocity distribution for a curated sample of 18 pulsars by explicitly incorporating the observational constraint of spin–velocity alignment. Using a hierarchical Bayesian framework that accounts for measurement uncertainties, we compare nine candidate velocity distribution models. We find that a Gamma distribution provides an adequate description of the inferred 3D velocities; however, the modest Bayes factor (1.65 relative to a single Maxwellian) indicates that the current data lack sufficient resolving power to discriminate decisively among the models considered.
The Gamma model is characterized by a peak velocity of $237^{+67}_{-84}$ $\text{km}\,\text{s}^{-1}$.
The reconstructed 3D velocities under alignment are systematically lower than those inferred under isotropy, indicating that projection effects can bias individual kick estimates high, while leaving the overall population scale largely unchanged within uncertainties. A complementary analysis of 465 pulsars with transverse velocity estimates favors a Log-Normal distribution for the full sample, while isolated young pulsars remain consistent with a Gamma-like profile. Our results underscore the importance of geometric assumptions in population inference and highlight the need for larger samples with improved distance and spin-axis measurements to place tighter constraints on natal kick physics.

\end{abstract}


\section{Introduction}\label{sec:1}

Pulsars---rapidly rotating magnetized neutron stars---serve as unique astrophysical laboratories. Since their serendipitous discovery \citep{1968Natur.217..709H}, they have provided profound insights into fundamental physics, stellar evolution, and the dynamics of the Galaxy. A defining characteristic of pulsars is their high space velocity, often exceeding that of their progenitor massive stars \citep{1994Natur.369..127L}. These “natal kicks”, imparted during the supernova explosion, are a direct probe of the violent, asymmetric processes at the heart of core collapse \citep{2012ARNPS..62..407J}. Quantifying the distribution of these kick velocities is essential, as it influences predictions for gravitational wave source populations \citep{2017ApJ...850L..40A}, binary evolution pathways \citep{2017ApJ...846..170T,2018MNRAS.479.4391M}, and the spatial distribution of neutron stars in the Milky Way \citep{1998ApJ...505..315C,2005MNRAS.360..974H}.

The origin of pulsar velocities is widely attributed to asymmetries in supernova explosions. Leading models for impulsive natal kicks include hydrodynamical instabilities \citep{2021Natur.589...29B}, asymmetric neutrino emission \citep{1998PhRvL..80.3694H}. In addition, the electromagnetic ``rocket" mechanism provides continuous acceleration over Myr timescales \citep{1975ApJ...201..447H,2018ApJ...865...61G,2022ApJ...931..123L,2023MNRAS.522.5879A}, potentially producing kicks from $\sim 200$ $\text{km}\,\text{s}^{-1}$ up to 1000 $\text{km}\,\text{s}^{-1}$ for extreme conditions. Conversely, electron-capture supernovae are typically associated with kicks $\lesssim$ 50 $\text{km}\,\text{s}^{-1}$ \citep{2013ApJ...771...28T,2017PASA...34...56D}, whereas ultra-stripped supernovae may yield effectively negligible natal kicks \citep{2025arXiv250508857V}. These mechanisms diverge not only in the predicted kick magnitude but also in the geometric relationship between the kick direction and the neutron star’s spin axis. For example, models like the electromagnetic rocket or anisotropic neutrino emission along the spin axis predict spin–kick alignment \citep{2020MNRAS.494.4665P}, whereas mechanisms involving multiple asymmetric shocks may produce kicks preferentially perpendicular to the spin \citep{1998Natur.393..139S}.

Progress in constraining the 3D pulsar velocity distribution has been hampered by a fundamental observational limitation: the radial velocity of pulsars is notoriously difficult to measure directly. Consequently, most studies have resorted to statistical methods, analyzing the distribution of transverse velocities under the simplifying assumption that the underlying 3D velocity field is isotropic \citep{2005MNRAS.360..974H,2006ApJ...643..332F}. Results from this approach remain inconclusive, with evidence supporting a single Maxwellian \citep{2005MNRAS.360..974H}, a bimodal Maxwellian \citep{2017A&A...608A..57V,2020MNRAS.494.3663I}, or a Log-Normal distribution \citep{disberg2025kinematically,2025ApJ...989L...8D}. This lack of consensus underscores the significant uncertainty in the inference of pulsar velocity distribution.

While the isotropic velocity prior is convenient, its physical validity is not established. Observational clues hint at potential deviations. For instance, some multi-wavelength studies of pulsar wind nebulae and radio polarization measurements have reported small projected misalignment angles ($\lesssim 10^{\circ}-15^{\circ}$) between spin and velocity vectors for individual sources \citep{2007ApJ...660.1357N,2017ApJ...844...84H,2018ApJ...856...18K}. More definitively, the first direct measurement of a normal pulsar's 3D velocity (PSR J0538+2817) revealed close spin-kick alignment in 3D space \citep{2021NatAs...5..788Y}. If such alignment is common, assuming isotropy could bias the inferred 3D velocity distribution \citep{2023ApJ...944..153M}. The current lack of consensus on the velocity model may therefore reflect not only statistical limitations but also an incorrect or overly restrictive geometric constraint. A more robust approach requires moving beyond a fixed isotropic framework to explicitly test how the inferred population properties depend on assumptions about spin-velocity geometry.

In this work, we present a Bayesian reconstruction of the pulsar 3D velocity distribution, explicitly incorporating the prior constraint of spin-velocity alignment and systematically comparing it against the traditional isotropic assumption. By combining updated polarization geometry, precise proper motions, and distances for a carefully selected sample of 18 pulsars, we derive the intrinsic velocity distribution and quantify how the results depend on the isotropy prior. We systematically compare a wide range of parametric models (Maxwellian, bimodal Maxwellian, Log-Normal, etc.) to determine which best describes the data. To provide broader context and validate our methodology, we also present in Appendix \ref{sec:append} a complementary hierarchical Bayesian analysis of the full pulsar catalog, analyzing young and recycled sub-populations under the isotropic framework. This comparative approach allows us to robustly assess the impact of geometric priors on the inferred population properties and to place our specific, geometry-constrained findings within the landscape of large-sample studies.

The structure of this paper is as follows. In Section \ref{sec:2}, we detail the selection of our pulsar sample and the processing of kinematic and polarization data. Section \ref{sec:3} presents the geometric framework linking the pulsar's 3D velocity to its observed polarization and proper motion. Our Bayesian methodology and the suite of tested velocity distribution models are introduced in Section \ref{sec:4}. The results of our analysis under both alignment and isotropic priors are presented in Section \ref{sec:5}. We discuss the astrophysical implications of our findings, compare them with previous work, and outline future prospects in Section \ref{sec:6}.

\section{Sample Selection}\label{sec:2}

This study focuses on the hypothesis of alignment between pulsar spin axes and velocity vectors. We aim to explore the 3D velocity distribution of pulsars under this geometric constraint through Bayesian analysis. This section details our sample selection process and criteria, and describes the sources and characteristics of the data we ultimately include in the analysis.

Our initial sample is compiled from multiple literature sources that provide the projected position angles of both the spin axis and the proper motion vector on the plane of the sky for individual pulsars. We begin with 82 pulsars gathered from references such as \citet{2005MNRAS.364.1397J,2006ApJ...639.1007W,2007ApJ...664..443R,2007MNRAS.381.1625J,2012MNRAS.423.2736N,2015MNRAS.453.4485F,2022ApJ...939...75Y}.

\begin{figure}[h]
    \centering
         \includegraphics[scale=0.35]{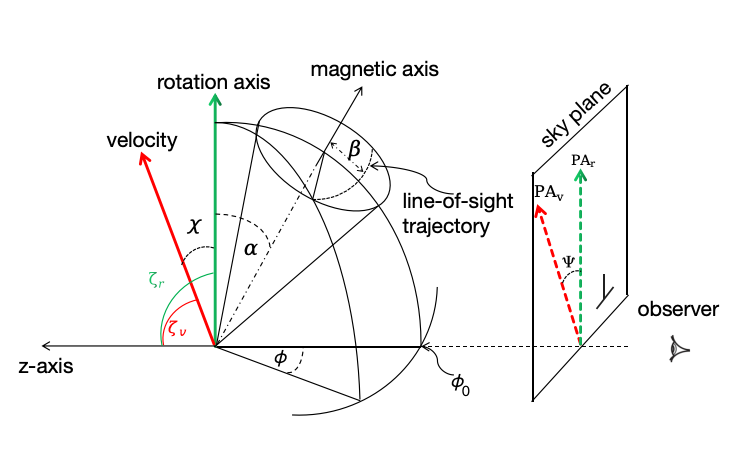}
         \caption{The 3D geometry of a pulsar's rotation axis, proper motion vector, and magnetic axis. The green solid line represents the pulsar's rotation axis vector, while the red solid line denotes the pulsar's 3D velocity vector. The angle $\chi$ signifies the 3D angular separation between these two vectors. The angles $\zeta_{r}$ and $\zeta_{v}$ represent the angular separations of the pulsar's rotation axis and velocity vector from the $z$-axis, respectively. The rotation axis (PA$\rm _r$) and proper motion vector (PA$\rm _v$) projected onto the plane of the sky are represented by the green and red dashed lines, respectively, with an angular separation of $\Psi$. Note that PA$_{r}$ is equivalent to PA$_0$ used elsewhere in this paper.}
         \label{image_1}
\end{figure}

Our goal is to derive the 3D velocity components under the condition where the velocity vector is aligned with the rotational axis. To infer this 3D alignment from observations, we first require that the projected directions of the spin axis and proper motion on the sky are themselves closely aligned. As Figure \ref{image_1} illustrates, this 2D alignment (a small value of $\Psi$) is a necessary, though not sufficient, condition for 3D alignment.

We account for measurement uncertainties in the polarization position angles. We adopt a threshold of $\left | \Psi \right |=\left | {\rm PA}_{\rm v} - {\rm PA}_0 \right | \lesssim 15^{\circ} $ for the velocity–rotational axis alignment criterion. Here, ${\rm PA}_0$ represents the projected position angle of the spin axis on the 2D sky plane at a fiducial phase $\phi=0$, corresponding to the PA$_{r}$
direction (measured clockwise) shown in Figure \ref{image_1}. Due to the well-known orthogonal polarization modes in pulsars \citep{2015MNRAS.453.4485F}, the observed position angle can exhibit a $90^\circ$ ambiguity. For PSRs J0014+4746 and J2257+5909, where the measured projected angle $|\Psi| \approx 90^\circ$, we assume the observed emission represents the orthogonal mode. Therefore, these sources are included in the samples. 

Applying this criterion yields an initial subset of 23 candidates. To preserve sensitivity to natal kick geometry and minimize the effects of long-term dynamical evolution, we exclude five pulsars from this group. Such exclusion is necessary because primordial spin–kick alignment can degrade as the pulsar velocity vector evolves in the Galactic potential \citep{2025PASA...42..106B}. Specifically, one source with an inferred velocity exceeding 3000 $\text{km}\,\text{s}^{-1}$ is removed, while four others are discarded because their characteristic ages ($\tau_c > 30$ Myr) imply substantial orbital evolution, leading to velocity vector deflections well in excess of 15$^{\circ}$—comparable to our adopted 2D alignment threshold.

\begin{figure}[h]
    \centering
         \includegraphics[scale=0.40]{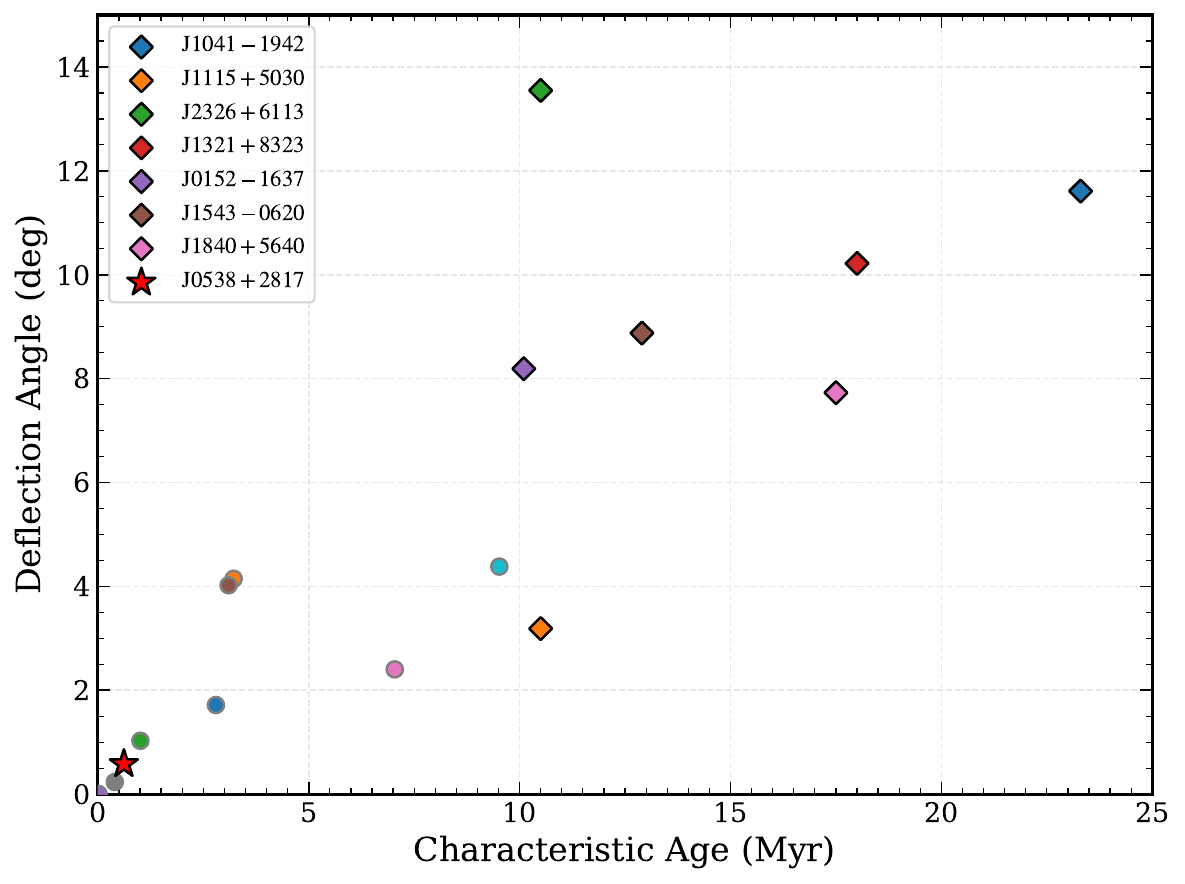}
         \caption{Velocity vector deflection angle induced by the Galactic potential as a function of characteristic age. Diamonds denote older sources ($\tau_c > 10$ Myr) for which dynamical backtracking is applied, while circles represent younger pulsars. Several young sources cluster near the coordinate origin because their small characteristic ages correspond to negligible deflection. The red star marks PSR J0538+2817.}
         \label{image_1_1}
\end{figure}

The resulting final sample consists of 18 sources, as listed in Table~\ref{tab1}. For seven pulsars with characteristic ages between 10 and 25 Myr whose cumulative deflection remains below 15$^{\circ}$ (see Figure~\ref{image_1_1}), we employ the \texttt{galpy} package \citep{2015ApJS..216...29B} to perform dynamical backtracking within a Milky Way potential model. This correction accounts for acceleration in the Galactic potential and provides an estimate of their birth velocities. Notably, PSR J0538+2817 is currently the only pulsar in our sample with a directly measured 3D velocity; for this source, we adopt the measurement from \citet{2021NatAs...5..788Y}.

Table \ref{tab1} presents the key observational and derived parameters for our final sample. For each source, we list the adopted distance. We update the distances for eight pulsars (J0323+3944, J1321+8323, J0152–1637, J1543–0620, J1840+5640, J1932+1059, J2157+4017) based on recent precise parallax measurements\footnote{\label{fn:1} \url{https://hosting.astro.cornell.edu/research/parallax/}} from VLBI astrometry. For the remaining pulsars, we take distances from the ATNF catalog, assuming a uniform 20\% uncertainty \citep{2017ApJ...835...29Y}. We also source characteristic ages ($\tau_{c}$) from the ATNF catalog\footnote{\label{fn:2}\url{https://www.atnf.csiro.au/research/pulsar/psrcat/}}.

For proper motion data, we compare values from the literature with those in the ATNF catalog, adopting the more recent measurement as the standard. We calculate transverse velocities following Equation (1) of \cite{2020MNRAS.494.3663I}. The table also lists the projected position angles of the proper motion vector (PA$_{\rm v}$) and the spin axis (PA$_{0}$), whose absolute difference $|\Psi|$ defines our 2D alignment criterion. The final column cites the primary literature source for the inclination angle $\zeta$ between the line of sight and the spin axis, as well as for PA$_{\rm v}$ and PA$_{0}$. For four sources marked with a superscript `$d$' in Table \ref{tab1}, $\zeta$ is derived from X-ray torus fitting \citep{2004ApJ...601..479N}.

\begin{deluxetable*}{ccccccccc}
\tablenum{1}
\tablecaption{Information of 18 pulsars used in our primary analysis.}
\tablewidth{0pt}
\setlength{\tabcolsep}{3pt} 
\renewcommand{\arraystretch}{1.1} 
\tablehead{
\colhead{Pulsar name} & \colhead{Distance} & \colhead{Age}& \colhead{$\mu_{\alpha}$} & \colhead{$\mu_{\beta}$} & \colhead{$\zeta$} & \colhead{PA$_{\rm v}$} & \colhead{PA$_{0}$} & \colhead{Ref} \\
\colhead{} & \colhead{(kpc)} &\colhead{(yr)}& \colhead{(mas/yr)} & \colhead{(mas/yr)} &\colhead{(deg)}& \colhead{(deg)} & \colhead{(deg)} & \colhead{}}
\startdata
J0908$-$1739 & 0.805$^{a}$ & $9.52 \times 10^{6}$ & $27 \pm 11$ & $-40 \pm 11$	& 80.4 & $167 \pm 2$ & $-27 \pm 14$ & 1 \\
J1041$-$1942 & 2.532$^{a}$ & $2.33 \times 10^{7}$ & $-1 \pm 3$ & $14\pm 5$ & 150.3 & $-4 \pm 2$ & $-12 \pm 1$ & 1 \\ 
J1115+5030 & $0.923^{a}$ & $1.05 \times 10^{7}$ & $22 \pm 3$ & $-51 \pm 3$	& 148.4 & $157 \pm 2$ & $-35 \pm 2$ & 1 \\
J1328$-$6038 & $1.424^{a}$ & $2.8 \times 10^{6}$ & $3 \pm 7$ & $54 \pm 2.3$ & 32.6 & $3 \pm 5$ & $-4 \pm 7$ & 1 \\
J1709$-$4429 & $2.6^{a}$ & $1.74 \times 10^{4}$ & $13 \pm 2$ & $-1 \pm 2$ & $55 \pm 0.2^{d}$ & $160 \pm 10$ & $163.6 \pm 7$ & 2 \\
J1913$-$0440 & $4.041^{a}$ & $3.22 \times 10^{6}$ & $7 \pm 13$ & $-5 \pm 9$ & 35.2 & $166 \pm 6$ & $-20 \pm 1$ & 1 \\
J2257+5909 & $3^{a}$ & $1.01 \times 10^{6}$ & $18 \pm 4.53$ & $-2 \pm 4.52$ & 125.9 & $106 \pm 12$ & $24 \pm 3$ & 1 \\
J2326+6113 & $2.733^{a}$ & $1.05 \times 10^{7}$ & $17 \pm 5$ & $-9 \pm 5$ & 23.2 & $63 \pm 24$ & $49 \pm 1$ & 1 \\
J0139+5814 & $2.703^{+0.328^{b}}_{-0.264}$ & $4.03 \times 10^{5}$ & $-19.11 \pm 0.07$ & $-16.6 \pm 0.07$ & 73.1 & $-131 \pm 1$ & $43 \pm 3$ & 1 \\
J0152$-$1637 & $2.273^{+1.573^{b}}_{-0.734}$ & $1.01 \times 10^{7}$ & $0.8 \pm 0.2$ & $-31.3 \pm 0.4$ & 88.7 & $173$ & $-86 \pm 9$ & 1 \\
J0534+2200 & $1.877^{+0.241^{b}}_{-0.192}$ & $1.26 \times 10^{3}$ & $-11.34 \pm 0.06$ & $2.65 \pm 0.14$ & $61.3\pm 0.01^{d}$ & $292 \pm 10$ & $124 \pm 1$ & 2 \\
J0538+2817 & $1.389^{+0.198^{b}}_{-0.199}$ & $6.18 \times 10^{5}$ & $-24.4 \pm 0.1$ & $57.2 \pm 0.1$ & $118.5 \pm 6.3$ & $337.0 \pm 0.1$ & $-18.5 \pm 1.7$ & 3 \\ 
J0835$-$4510 & $0.286^{+0.017^{b}}_{-0.015}$ & $1.13 \times 10^{4}$ & $-49.68 \pm 0.06$ & $29.9 \pm 0.1$ & $63.6 \pm 0.1^{d}$ & $301 \pm 2$ & $35 \pm 10$ & 4 \\
J1321+8323 & $1.031^{+0.174^{b}}_{-0.041}$ & $1.8 \times 10^{7}$ & $-52.66 \pm 0.09$ & $32.45 \pm 0.13$ & 162.6 & $-76 \pm 14$ & $26 \pm 3$ & 1 \\
J1543$-$0620 & $3.03^{+0.478^{b}}_{-0.237}$ & $1.29 \times 10^{7}$ & $-16.79 \pm 0.04$ & $-0.3 \pm 0.13$ & 39.6 & $-103 \pm 10$ & $66 \pm 2$ & 1 \\
J1840+5640 & $1.522^{+0.018^{b}}_{-0.137}$ & $1.75 \times 10^{7}$ & $-31.21 \pm 0.03$ & $-29.1 \pm 0.06$ & 30.6 & $-125 \pm 5$ & $56 \pm 2$ & 1 \\
J1932+1059 & $0.361^{+0.122^{b}}_{-0.073}$ & $3.1 \times 10^{6}$ & $94.09 \pm 0.11$ & $42.99 \pm0.16$ & 68 & $65 \pm 1$ & $78 \pm 1$ & 1 \\
J2157+4017 & $3.571^{+0.974^{b}}_{-0.630}$ & $7.04 \times 10^{6}$ & $16.13 \pm 0.1$ & $4.12 \pm0.12$ & 104.4 & $83 \pm 7$ & $-7 \pm 7$ & 1 \\
\enddata
\label{tab1}
\tablecomments{References for distance measurements listed in the second column are: data with a superscript $a$ is from ATNF data\footref{fn:2}, superscript $b$ for precise parallax measurements\footref{fn:1}, and superscript $c$ from \citet{2009ApJ...698..250C}. The age data in the fourth column are sourced from the ATNF data\footref{fn:2}. Inclination angle $\zeta$ marked with superscript $d$ is obtained from \citet{2004ApJ...601..479N}. All other data entries are from references listed in column 9: 1) \cite{2015MNRAS.453.4485F}, 2) \cite{2012MNRAS.423.2736N}, 3) \cite{2022ApJ...939...75Y}, 4) \cite{2006ApJ...639.1007W}.}
\end{deluxetable*}

\section{Geometric Framework for 3D Velocity Reconstruction}\label{sec:3}

To reconstruct the intrinsic three-dimensional (3D) velocities of the pulsar sample, we must relate the observed projected quantities to the true spatial vectors. As illustrated in Figure \ref{image_1}, we define the pulsar spin axis and the velocity vector in 3D space as
$\vec{S} =(S_{x},S_{y},S_{z})$ and $\vec{V} =(V_{x},V_{y},V_{z})$, respectively. Their projections onto the plane of the sky are denoted as $\vec{S}_{1}=(S_{x},S_{y},0)$ and $\vec{V}_{1}=(V_{x},V_{y},0)$ corresponding to the observed position angles PA$_{r}$ and PA$_{\rm v}$. The projected misalignment angle, $\Psi$, between these vectors on the sky plane is given by:
\begin{equation}
        \cos(\Psi)=\frac{\vec{S}_{1} \cdot \vec{V}_{1}}
{|\vec{S}_{1}||\vec{V_{1}} |}
=\frac{S_{x}V_{x}+S_{y}V_{y}}{|\vec{S}_{1} ||\vec{V}_{1} |}.
\end{equation}
The intrinsic 3D misalignment angle $\chi$ can be expressed as:
\begin{equation}
    \cos(\chi)=\frac{\vec{S} \cdot \vec{V}}
{|\vec{S}||\vec{V} |}
=\frac{S_{x}V_{x}+S_{y}V_{y}+S_{z}V_{z}}{\sqrt{S_{x}^{2}+S_{y}^{2}+S_{z}^{2}} \cdot 
\sqrt{V_{x}^{2}+V_{y}^{2}+V_{z}^{2}}  }.
\label{eq2}
\end{equation}
\begin{figure}[h]
    \centering
         \includegraphics[scale=0.45]{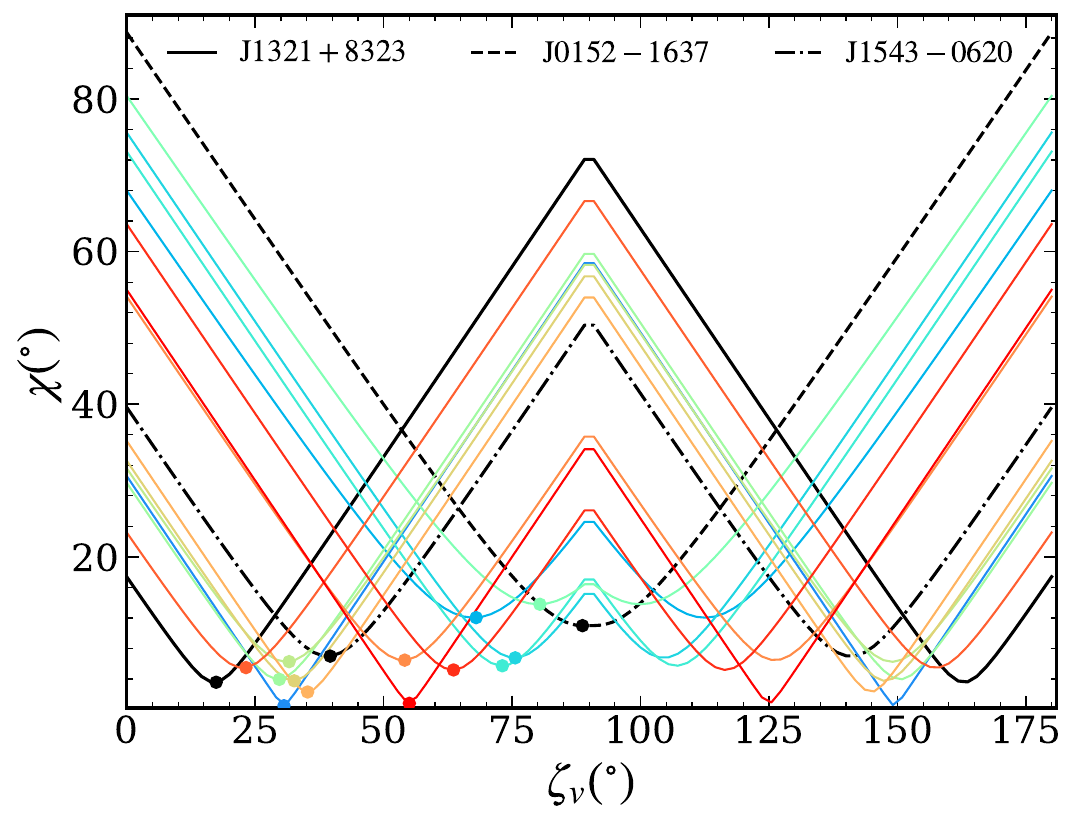}
         \caption{
The angle \(\chi\) between the pulsar spin axis and velocity vector as a function of \(\zeta_v\) (the angle between the 3D velocity vector and the  \(z\)-axis, the line-of-sight). As shown in Figure \ref{image_1}, the projected angle \(\Psi\) between the velocity and spin axis on the 2D sky plane remains fixed, while the lack of radial velocity measurements allows the 3D velocity vector to pivot along the line of sight, resulting in a degeneracy of \(\zeta_v\) spanning \(0^\circ\)–\(180^\circ\). Each pulsar is color-coded, with three representative sources explicitly labeled in the legend. The scatter points mark the $\chi$ values calculated using Equation \ref{eq6} when $\zeta_{v}=\zeta_{r}$ for individual sources.
}
         \label{image_2}
\end{figure}
Let $\zeta_{r}$ be the angle between $\vec{S}$ and the line of sight (z-axis), and $\zeta_{v}$ the angle between $\vec{V}$ and the z-axis. The inclination $\zeta=\zeta_{r}$ is obtained for each pulsar from rotating vector model (RVM) fits or X-ray torus modeling. Using spherical trigonometry, Equation \ref{eq2} can be rewritten as
\begin{equation}
    \small
    \begin{split}
    \cos(\chi) & =\frac{|\vec{S}||\vec{V}|(\sin \zeta_{r}\cdot \sin \zeta_{v}\cdot \cos(\Psi) +\cos \zeta_{r} \cdot cos \zeta_{v})}{|\vec{S}||\vec{V}|} \\
    &=\sin \zeta_{r} \cdot \sin \zeta_{v} \cdot \cos(\Psi) + \cos \zeta_{r} \cdot \cos \zeta_{v}. 
    \label{eq6}
    \end{split}
\end{equation}

While $\zeta_{r}$ is observationally constrained, $\zeta_{v}$ remains unknown for most pulsars due to the lack of radial velocity measurements. Determining the true velocity magnitude, $V_{3D}=V_{\perp}/\sin\zeta_{v}$, therefore requires constraining $\zeta_{v}$. 
In this study, we resolve this degeneracy by assuming the spin-velocity alignment. Under the assumption that the kick is nearly aligned with the spin axis in 3D space, the misalignment angle $\chi$ must be minimized.

We determine the velocity inclination $\zeta_{v}$ for each source by minimizing $\chi$ given the observed $\zeta_{r}$ and $\Psi$. As shown in Figure \ref{image_2}, which plots $\chi$ as a function of $\zeta_{v}$, the global minimum for $\chi$ occurs at $\zeta_{v} \approx \zeta_{r}$. Since our sample selection already enforces approximate 2D alignment (small $\Psi$), setting $\zeta_{v}=\zeta_{r}$ effectively aligns the vectors in 3D space. This geometric constraint allows us to recover the full 3D velocity vector.

Using this framework, we calculate $V_{3D}$ for 22 pulsars in our sample. For PSR J0538+2817, we adopt the directly measured 3D velocity from \cite{2021NatAs...5..788Y}. It is worth noting that while the solution $\zeta_{v}=180^{\circ}-\zeta_{r}$ yields an identical velocity magnitude, it implies a velocity directed away from the observer rather than towards; however, as we are interested in the magnitude distribution, this directional ambiguity does not affect the subsequent statistical analysis.

\section{Methods}\label{sec:4}
\subsection{Hierarchical Bayesian inference}

We infer the intrinsic 3D velocity distribution of pulsars using hierarchical Bayesian inference \citep[e.g.,][]{Thrane:2018qnx,You:2024bmk}. For each pulsar $i = 1,\dots,N$, we denote the latent 3D velocity magnitude by $v_i$ and define the measurement vector
\begin{equation}
\vec d_i =
\{\mu_{\alpha,i}, \mu_{\delta,i}, D_i, \zeta_i, PA_{v,i}, PA_{0,i}\},
\end{equation}
where $\mu_{\alpha,i}$ and $\mu_{\delta,i}$ are the proper-motion components, $D_i$ is the distance, $\zeta_i$ is the spin-axis inclination angle, and $PA_{v,i}$ and $PA_{0,i}$ are the projected position angles of the velocity and spin axis, respectively. Measurement uncertainties are assumed to be independent and Gaussian.
All quantities are taken from Table~\ref{tab1} and Section~\ref{sec:2}.

For a population model $M$ with hyperparameters $\Lambda$, intrinsic velocities are assumed to be drawn from a parametric distribution $\pi(v_i|\Lambda,M)$. The hyperprior $\pi(\Lambda|M)$ encodes prior assumptions on the population parameters (e.g., uniform priors over physically motivated ranges; see Section~\ref{sec:dist-models}).

The hierarchical likelihood is
\begin{equation}
\mathcal{L}(\{\vec d\}|\Lambda,M)
=
\prod_{i=1}^{N}
\int
L(\vec d_i|v_i)\,
\pi(v_i|\Lambda,M)\,
dv_i ,
\label{eq:hier_likelihood}
\end{equation}
where $L(\vec d_i|v_i)$ is the single-pulsar likelihood.

Following the posterior-reweighting formalism, we do not evaluate $L(\vec d_i|v_i)$ analytically. Instead, for each pulsar we generate posterior samples of $v_i$ by Monte Carlo propagation of the Gaussian uncertainties in $\vec d_i$, incorporating the sample selection and dynamical corrections described in Section~\ref{sec:2} together with the geometric relations developed in Section~\ref{sec:3}. This yields a posterior distribution $p(v_i|\vec d_i)$, obtained under a reference prior $\pi_0(v_i)$, taken to be uniform in $v_i$ over the sampled range.

By Bayes’ theorem,
\begin{equation}
p(v_i|\vec d_i)
=
\frac{L(\vec d_i|v_i)\,\pi_0(v_i)}{Z_{0,i}},
\end{equation}
where $Z_{0,i}$ is the single-source evidence. Rearranging,
\begin{equation}
L(\vec d_i|v_i)
=
Z_{0,i}
\frac{p(v_i|\vec d_i)}{\pi_0(v_i)}.
\label{eq:likelihood_reweight}
\end{equation}

Substituting Eq.~(\ref{eq:likelihood_reweight}) into Eq.~(\ref{eq:hier_likelihood}) gives
\begin{equation}
\mathcal{L}(\{\vec d\}|\Lambda,M)
=
\prod_{i=1}^{N}
Z_{0,i}
\int
p(v_i|\vec d_i)
\frac{\pi(v_i|\Lambda,M)}{\pi_0(v_i)}
\, dv_i .
\end{equation}

For each pulsar $i$, let $\{v_{i,k}\}_{k=1}^{n_i}$ denote the $n_i$ posterior samples drawn from $p(v_i|\vec d_i)$. The integral is evaluated via Monte Carlo summation,
\begin{equation}
\int
p(v_i|\vec d_i)
\frac{\pi(v_i|\Lambda,M)}{\pi_0(v_i)}
\, dv_i
\approx
\frac{1}{n_i}
\sum_{k=1}^{n_i}
\frac{\pi(v_{i,k}|\Lambda,M)}{\pi_0(v_{i,k})},
\end{equation}
where $n_i$ is the number of posterior samples for pulsar $i$.
This estimator is unbiased in the limit of large $n_i$.
Therefore, the hierarchical likelihood becomes
\begin{equation}
\mathcal{L}(\{\vec d\}|\Lambda,M)=\prod_{i=1}^N \frac{Z_{0,i}}{n_i} \sum_{k=1}^{n_i} \frac{\pi\left(v_{i, k} \mid \Lambda, M\right)}{\pi_0\left(v_{i, k}\right)}.
\end{equation}
Since $Z_{0,i}$ does not depend on $\Lambda$, it cancels in evidence ratios.

The hyperposterior is
\begin{equation}
p(\Lambda|\{\vec d\},M)
\propto
\mathcal{L}(\{\vec d\}|\Lambda,M)\,
\pi(\Lambda|M),
\end{equation}
and the model evidence is
\begin{equation}
Z_\Lambda(M)
=
\int
\mathcal{L}(\{\vec d\}|\Lambda,M)\,
\pi(\Lambda|M)\, d\Lambda .
\end{equation}

Bayes factors between two models $M_1$ and $M_2$ are defined as
\begin{equation}
\mathrm{BF}_{12}
=
\frac{Z_\Lambda(M_1)}{Z_\Lambda(M_2)}.
\end{equation}

\subsection{Distribution models of pulsar velocity}
\label{sec:dist-models}

In this section, we select 9 statistical distribution models for comparison. In our subsequent analysis, we use the software package BILBY \citep{2019ApJS..241...27A} for parameter estimation and evidence calculation. According to Bayesian methodology, a Log-Uniform (LU) prior is generally more appropriate when the parameter of interest spans several orders of magnitude.
The LU distribution is defined as
\begin{equation}
    \pi(v|\Lambda \left \{ v_{\rm min},v_{\rm max} \right \}) 
     = \left\{\begin{matrix}
  \frac{1}{v \ln(v_{\rm max}/v_{\rm min})} & v_{\rm min}\le v \le v_{\rm max}\\
  0& \text{otherwise,}
\end{matrix}\right.
\end{equation}
where $v_{\rm min}$ and $v_{\rm max}$ are the minimum and maximum pulsar velocity, respectively. 

We have also selected the Log-Normal (LN) distribution, defined as:
\begin{equation}
        \pi(v|\Lambda \left \{\mu,\sigma \right \})
        =\frac{1}{v\sigma \sqrt{2\pi}} \text{exp} \left [ -\frac{(\ln v-\mu)^{2}}{2 \sigma^{2}}  \right ]. 
\end{equation}
The Gamma (GA) distribution is given by 
\begin{equation}
        \pi(v|\Lambda \left \{k,\theta \right \}) 
        = \frac{v^{(k-1)} \text{exp} (- \frac{v}{\theta})}{\theta^{k}\Gamma(k)}.
\end{equation}
Here, $\Gamma(k)$ denotes the Gamma function. The probability distribution of Gaussian (G) model is defined as 
\begin{equation}
    \pi(v|\Lambda \left \{ \mu,\sigma \right \})
    =\frac{1}{\sigma \sqrt{2\pi}} \text{exp} \left [ -\frac{1}{2} \left ( \frac{v-\mu}{\sigma}  \right )^{2}  \right ] ,
\end{equation}
where $\sigma$ and $\mu$ are the standard deviation and mean value, respectively. When using the Gaussian model we treat it as a truncated Gaussian on $v\ge 0$. We also consider the two-Gaussian (2G) model. For the 2G model, $\Lambda=\left \{\mu_{1},\sigma_{1},\mu_{2},\sigma_{2},\alpha \right \}$, where $\alpha$ is the weight ratio between the two Gaussian components.
Another, Maxwellian (M) distribution is given by:
\begin{equation}
        \pi(v|\Lambda \left \{ \sigma \right \})
        =\left\{\begin{matrix}
  \frac{\sqrt{\frac{2}{\pi} }v^{2} e^{-v^{2}/(2\sigma^{2})}}{\sigma^{3}} & ,v>0\\
  0&,\text{otherwise.}
\end{matrix}\right.
\end{equation}
We also have $\Lambda=\left \{\sigma_{1} ,\sigma_{2}, \alpha\right \}$ for bimodal Maxwell (2M) distribution, here $\alpha$ is the weight ratio between two Maxwell components.
For the Exponential (E) model, the probability distribution is 
\begin{equation}
        \pi(v|\Lambda \left \{ \lambda \right \})=\left\{\begin{matrix}\lambda e^{-\lambda v} & v\ge 0\\0&,\text{otherwise.}
        \end{matrix}\right.
\end{equation}
The skewed Student's t (SST) distribution is given by 
\begin{equation}
    \pi(v|\Lambda \left \{ \mu,\sigma,\nu,\tau \right \})=
\left
\{\begin{matrix}
  \frac{c}{\sigma_{0}}\left [ 1+\frac{\nu^{2}z^{2}}{\tau}  \right ]^{-(\tau+1)/2}  & v<\mu_{0} \\
  \frac{c}{\sigma_{0}}\left [ 1+\frac{z^{2}}{\nu^{2} \tau}  \right ]^{-(\tau+1)/2}  & v\ge \mu_{0},
\end{matrix}\right.
\end{equation}
where $\mu $ and $\sigma$ are the mean value and standard deviation of the corresponding Gaussian distribution, respectively. $\nu$ and $\tau$ are used to describe the degree of skewness. $\mu_{0}=\mu-\sigma\kappa /s$, $\sigma_{0}=\sigma/s$, $z=(v-\mu_{0})/\sigma_{0}$, $c= 2\nu [(1+\nu^{2})B(1/2,\tau/2)\tau^{1/2}]^{-1}$, $\kappa=2\tau^{1/2}(\nu-\nu^{-1})[(\tau-1)B(1/2,\tau/2)]^{-1}$, $s=\left [ \frac{\tau(\nu^{2}+\nu^{-2}-1)}{\tau-2}-\kappa^{2} \right ]^{1/2}$, where B is the beta function.

In the selection of parameter prior intervals, we set the prior range of the $\mu$ parameter of the Gaussian, bimodal-Gaussian, and SST distributions to [10,4000] $\text{km}\,\text{s}^{-1}$, and set the $\sigma$ priors to [10,1500] $\text{km}\,\text{s}^{-1}$. For the Maxwellian distribution, whose mean is mathematically linked to its parameter through $\mu= \sigma \sqrt{\frac{8}{\pi}}$, the corresponding $\sigma$ prior interval is adjusted to [10,2500] $\text{km}\,\text{s}^{-1}$. Regarding the shape parameters k for the Gamma distribution and $\sigma$ for the Log-Normal distribution, both were consistently assigned the interval [0.01,10]. The minimum velocity parameter $v_{\rm min} $ and the maximum velocity parameter $v_{\rm max}$ were respectively restricted to [0.001,400] and [400,4000] in all applicable models. A comprehensive summary of the prior interval selections for the 9 distribution models is provided in Table \ref{tab2}. 

In Table \ref{tab2} and Table \ref{tab4}, “U” denotes a uniform prior and “logU” a log-uniform prior.
As a practical criterion, parameters spanning three or more orders of
magnitude are assigned log-uniform priors, whereas narrower ranges
are modeled with uniform priors. For brevity, units are omitted in Tables \ref{tab2} and \ref{tab4}; all velocity-scale parameters are reported in $\text{km}\,\text{s}^{-1}$.

\begin{figure}[h]
    \centering
         \includegraphics[scale=0.4]{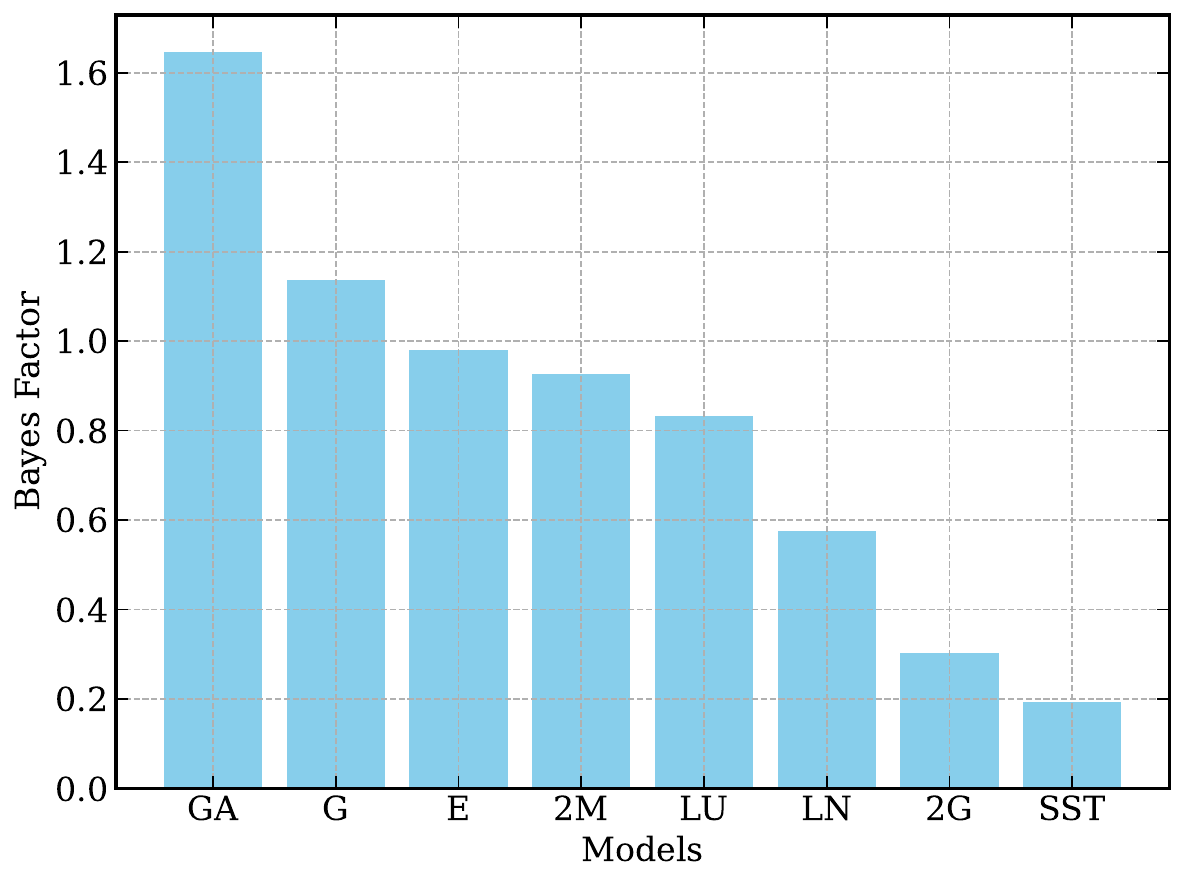}
         \caption{Bayesian factors of 8 distribution models in comparison to the Maxwellian model for the pulsar 3D velocity distributions assuming spin-velocity alignment.}
         \label{image_3}
\end{figure}

\begin{table*}[ht!]
\caption{Prior, 1$\sigma$ posterior intervals and BF of each model relative to the M model for 8 velocity-distribution models for the spin-velocity alignment and isotropy cases.}
\centering
\setlength{\tabcolsep}{4pt} 
\renewcommand{\arraystretch}{1.1} 
\begin{tabular*}{\textwidth}{@{\extracolsep{\fill}}cccccccc}
\hline\hline
\multirow{2}{*}{Model} &
\multirow{2}{*}{Parameters} &
\multicolumn{3}{c}{Alignment case } &
\multicolumn{3}{c}{Isotropy case }\\
\cline{3-8}
 & & Prior &Posterior&BF &  Prior&Posterior&BF\\
\hline
\multirow{2}{*}{LU} 
& $v_{\text{min}}$ & $\text{logU}[0.01,400]$ &$74^{+12}_{-20}$ & \multirow{2}{*}{0.83}&$\text{U}[1,400]$ &$187^{+53}_{-47}$ & \multirow{2}{*}{1.34}\\ 
& $v_{\text{max}}$ & $\text{U}[400,4000]$ &$1156^{+349}_{-154}$&\multirow{2}{*}{}& $\text{U}[400,10000]$ &$449^{+77}_{-37}$&\\ 
\hline 
\multirow{2}{*}{GA} 
& k & $\text{logU}[0.01,10]$ & $2.53^{+1.13}_{-0.81}$& \multirow{2}{*}{1.65}& $\text{logU}[0.01,50]$ & $9.06^{+11.58}_{-4.35}$& \multirow{2}{*}{1.55}\\ 
& $\theta$& $\text{U}[1,400]$ &$156^{+80}_{-51}$&\multirow{2}{*}{} & $\text{U}[1,400]$ &$33^{+32}_{-18}$& \\ 
\hline 
\multirow{2}{*}{LN}
& $\mu$ & $\text{U}[1,10]$ & $5.78^{+0.18}_{-0.19}$ & \multirow{2}{*}{0.57} & $\text{U}[1,10]$ & $5.68^{+0.09}_{-0.11}$ & \multirow{2}{*}{0.29}\\
& $\sigma$ &$\text{logU}[0.01,10]$ & $0.68^{+0.16}_{-0.12}$ &\multirow{2}{*}{}&$\text{U}[0.1,10]$ & $0.26^{+0.15}_{-0.11}$ &\\
\hline
\multirow{3}{*}{E}
& $v_{\text{min}}$ & $\text{logU}[0.01,400]$& $0.63^{+9.64}_{-0.59}$ &\multirow{3}{*}{0.98}&$\text{logU}[1,400]$& $45^{+30}_{-30}$ &\multirow{3}{*}{0.0005}  \\
& $v_{\text{max}}$ & $\text{U}[400,4000]$ & $2016^{+1187}_{-748}$ &\multirow{3}{*}{}&$\text{U}[400 ,10000]$ & $5286^{+3265}_{-3200}$ &\\
& $\lambda$ & $\text{logU}[0.0001,0.01]$ & $0.0026^{+0.0007}_{-0.0006}$ &\multirow{3}{*}{}& $\text{logU}[0.0001,0.01]$ & $0.0032^{+0.0009}_{-0.0007}$ &\\
\hline
\multirow{2}{*}{G}
& $\mu$ & $\text{U}[10,4000]$ &$257^{+128}_{-156}$ &\multirow{2}{*}{1.14}& $\text{U}[10,4000]$ &$300^{+27}_{-31}$ &\multirow{2}{*}{0.82}\\
& $\sigma$ & $\text{U}[10,1500]$ & $360^{+119}_{-91}$ & \multirow{2}{*}{}&$\text{U}[10,1500]$ & $80^{+40}_{-35}$ &\\
\hline
\multirow{5}{*}{2G}
& $\mu_{1}$ & $\text{U}[10,4000]$ & $221^{+134}_{-141}$ &\multirow{5}{*}{0.30}& $\text{U}[10,4000]$ & $293^{+33}_{-53}$ &\multirow{5}{*}{0.11}\\
& $\sigma_{1}$ & $\text{U}[10,1500]$ & $346^{+180}_{-126}$ &\multirow{5}{*}{}
& $\text{U}[10,1500]$ & $85^{+66}_{-40}$ &\\
& $\mu_{2}$ & $\text{U}[10,4000]$ & $626^{+1576}_{-303}$ &\multirow{5}{*}{}& $\text{U}[10,4000]$ & $1435^{+1703}_{-1084}$ &\\
& $\sigma_{2}$ &$\text{U}[10,1500]$ & $454^{+629}_{-260}$ &\multirow{5}{*}{} &$\text{U}[10,1500]$ & $599^{+609}_{-507}$ &\\
& $\alpha$ & $\text{U}[0.01,0.99]$ & $0.84^{+0.12}_{-0.47}$ &\multirow{5}{*}{}& $\text{U}[0.01,0.99]$ & $0.94^{+0.04}_{-0.26}$ &\\
\hline
\multirow{1}{*}{M}
& $\sigma $ & $\text{U}[10,2500]$ & $270^{+32}_{-28}$ &\multirow{1}{*}{1} & $\text{U}[10,2500]$ & $183^{+24}_{-22}$ &1\\
\hline
\multirow{3}{*}{2M}
& $\sigma_{1}$ & $\text{U}[10,2500]$ & $201^{+55}_{-73}$ &\multirow{3}{*}{0.93} & $\text{U}[10,2500]$ & $174^{+26}_{-28}$ &\multirow{3}{*}{0.68}\\
& $\sigma_{2}$ & $\text{U}[10,2500]$ & $396^{+417}_{-99}$ &\multirow{3}{*}{}& $\text{U}[10,2500]$ & $413^{+1322}_{-217}$ &\\
& $ \alpha$ & $ \text{U}[0.01,0.99]$ & $0.69^{+0.23}_{-0.41}$ &\multirow{3}{*}{}& $\text{U}[0.01,0.99]$ & $0.89^{+0.08}_{-0.55}$ &  \\
\hline
\multirow{4}{*}{SST}
& $\mu$ & $ \text{U}[10,4000]$ & $410^{+91}_{-68}$& \multirow{4}{*}{0.19}&$ \text{U}[10,4000]$ & $297^{+31}_{-33}$& \multirow{4}{*}{0.84}\\
& $\sigma$ & $\text{U}[10,1500]$ & $306^{+170}_{-72}$&\multirow{4}{*}{} & $\text{U}[10,1500]$ & $89^{+74}_{-40}$& \\
& $\nu$ & $\text{logU}[0.01,100]$ &$11^{+38}_{-8}$ & \multirow{4}{*}{}& $\text{logU}[0.01,100]$ &$0.18^{+3.45}_{-0.15}$  & \\
& $\tau$ & $\text{U}[2.01,30]$ &$7.75^{+11.50}_{-4.71}$ &\multirow{4}{*}{}& $\text{U}[2.01,30]$ &$6.20^{+11.94}_{-3.69}$ & \\
\hline
\end{tabular*}
\label{tab2}
\end{table*}

\begin{figure}[h]
    \centering
         \includegraphics[scale=0.6]{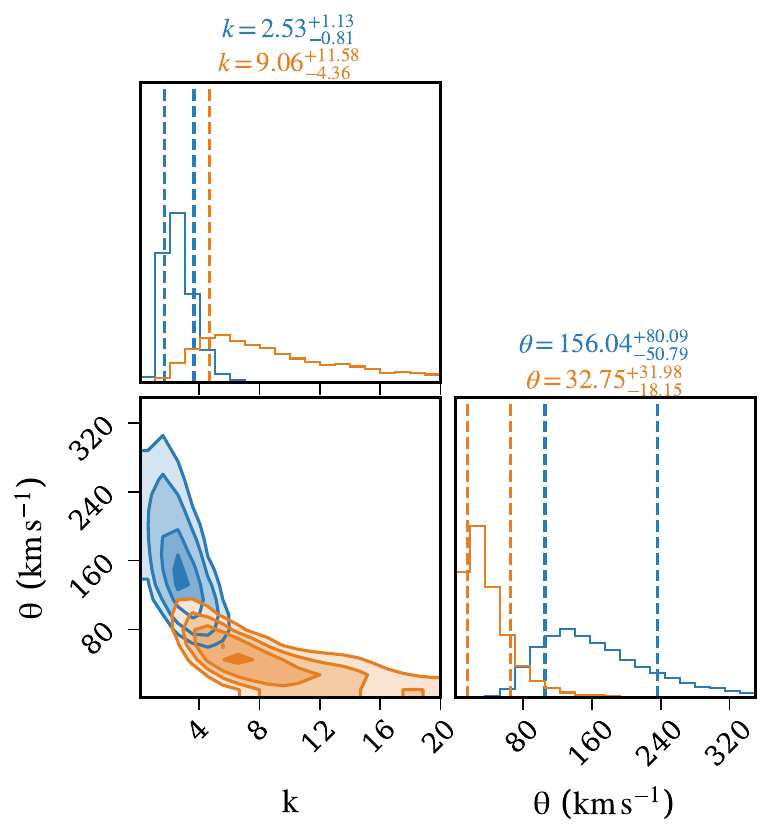}
         \caption{Posterior parameter distributions of the Gamma model. Blue and orange contours show results for spin-velocity alignment and isotropy prior, respectively.}
         \label{image_4}
\end{figure}

\begin{figure}[h]
    \centering
         \includegraphics[scale=0.45]{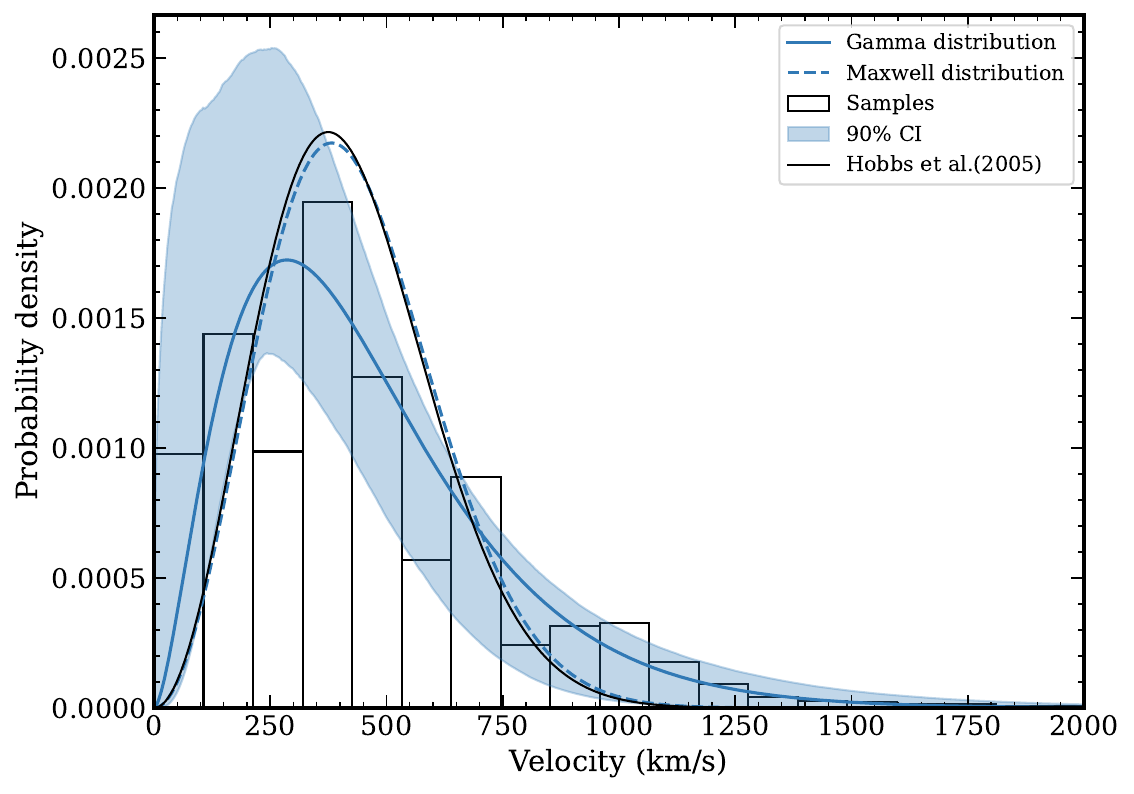}
         \caption{Pulsar 3D velocity distribution reconstructed assuming spin-velocity alignment. The black histogram represents the 3D velocity distributions of 18 pulsars. The blue solid (dashed) line shows the posterior predictive distribution of the Gamma (Maxwellian) model. The blue shaded region indicates the 90$\%$ confidence interval of the Gamma distribution, and the black curve displays the results from \citet{2005MNRAS.360..974H} as a reference.}
         \label{image_5}
\end{figure}

\section{Results}\label{sec:5}
\subsection{Velocity distribution under the alignment prior}

To identify which parametric form best describes the reconstructed 3D pulsar kick velocities under the spin–velocity alignment hypothesis, we apply a hierarchical Bayesian framework to nine candidate models. The models, prior ranges, and posterior estimates are summarized in Table \ref{tab2}. The corresponding Bayes factors for each model relative to the single Maxwellian model are shown in Figure \ref{image_3}.

Our analysis reveals that the GA model yields the highest evidence among the candidates, with a Bayes factor of 1.65 relative to the single Maxwellian. According to the classification scheme of \citet{kass1995bayes}, this level of support is not statistically significant and does not allow us to discriminate meaningfully between the models.

In Figure \ref{image_4}, blue contours show the posterior distribution for the parameters of the Gamma model under the alignment prior. The model is characterized by a shape parameter k=$2.53^{+1.13}_{-0.81}$ and a scale parameter $\theta=156^{+80}_{-51}$
$\text{km}\,\text{s}^{-1}$, resulting in a mean velocity of $k\theta=395^{+72}_{-60}$ $\text{km}\,\text{s}^{-1}$. This indicates that the pulsar kick distribution is well-represented by a single continuous Gamma component, with its peak located at $(k-1)\theta= 237^{+67}_{-84}$ $\text{km}\,\text{s}^{-1}$.

The single Maxwellian model, which remains to be a plausible model, yields a dispersion of $\sigma = 270^{+32}_{-28}$ $\text{km}\,\text{s}^{-1}$ (mean velocity $431^{+52}_{-44}~ \text{km}\,\text{s}^{-1}$).
This posterior is shown as a blue dashed line in Figure \ref{image_5}, which is consistent with the classic result (the black solid line) of \citet{2005MNRAS.360..974H}.

Our results do not favor or strongly disfavor other models.
The LN posterior parameters we obtain ($\mu=5.78^{+0.18}_{-0.19}$, $\sigma=0.68^{+0.16}_{-0.12}$) indicate a systematically lower median velocity and a narrower dispersion than those reported by \citet{disberg2025kinematically} under an isotropic prior. This shift highlights how the spin–alignment constraint modifies the inferred velocity distribution.

\subsection{Velocity distribution under the isotropic prior}

To assess the impact of our geometric assumption, we repeat the analysis after relaxing the alignment constraint. We instead draw the velocity inclination angle $\zeta_{v}$ uniformly in $\cos \zeta_{v}$ (i.e., assuming an isotropic 3D orientation) while keeping the observed transverse speed fixed. The resulting 3D speeds are then analyzed with the same Bayesian pipeline.
The results are summarized in the right half of Table \ref{tab2}. Under isotropy, the GA model again yields the highest evidence among the candidates (BF = 1.55 relative to the single Maxwellian), although this difference is statistically insignificant. The posterior parameters shift toward a higher shape parameter ($k=9.06^{+11.58}_{-4.35}$) and a lower scale parameter ($\theta=33^{+32}_{-18}$ $\text{km}\,\text{s}^{-1}$), as shown by the orange contours in Figure~\ref{image_4}. This parameter shift reflects a compensatory adjustment that preserves the overall velocity scale.
The resulting peak velocity ($265^{+35}_{-42}$ $\text{km}\,\text{s}^{-1}$) is statistically consistent with that inferred under the alignment assumption. Although the reconstructed 3D velocities of individual pulsars are systematically lower when alignment is imposed, the characteristic population-scale velocity remains broadly similar within uncertainties.

\subsection{Radial velocity distribution}

\begin{figure}[h]
    \centering
         \includegraphics[scale=0.46]{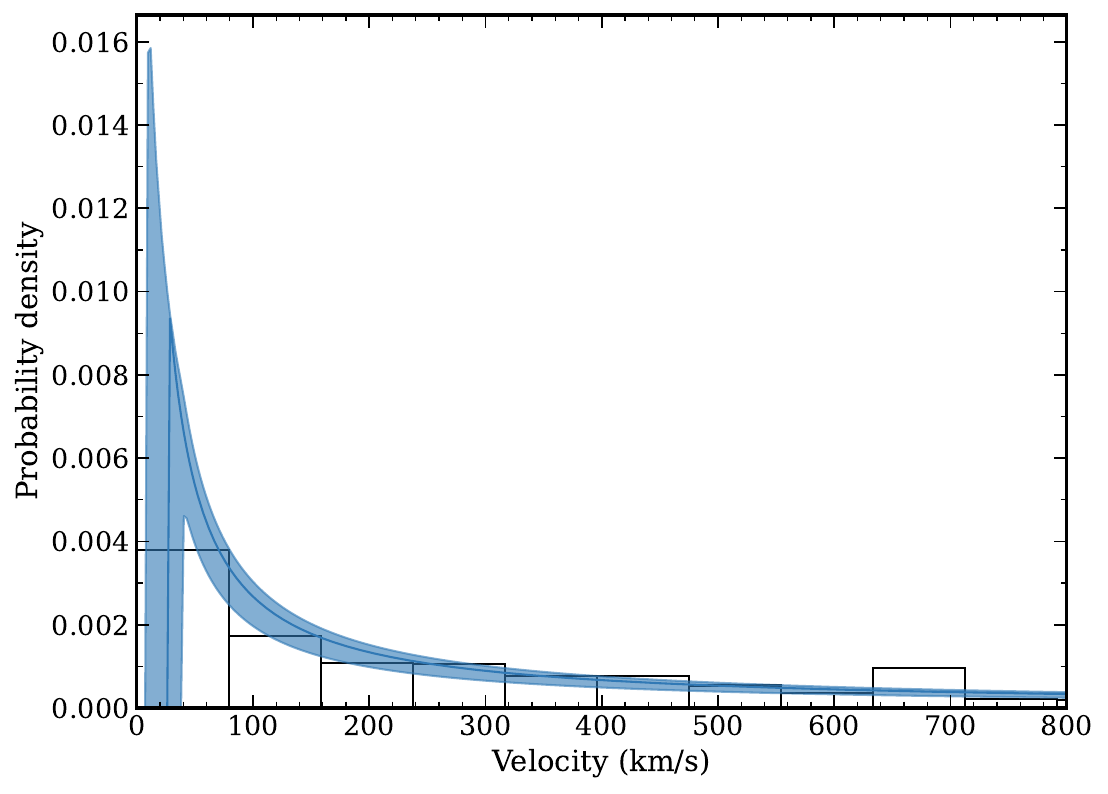}
         \caption{Pulsar radial velocity distribution: the histogram represents the distribution of 18 pulsars assuming spin-kick alignment, while the blue line and shaded region indicate the posterior predictive distribution and its 90$\%$ confidence interval based on the Log-Uniform model.}
         \label{image_6}
\end{figure}

We analyze the radial velocity component for our sample under the alignment assumption. Comparing the same nine models, we find that the Log-Uniform (LU) distribution yields the highest evidence for the radial velocity data, closely followed by the bimodal Maxwellian (2M) model, with a Bayes factor of 1.03 for LU relative to 2M. This negligible difference indicates that the two models are statistically indistinguishable given the current data. In contrast, the single Maxwellian is strongly disfavored, with a logarithmic Bayes factor of 15.32 relative to LU.
The posterior parameters for the LU model are $v_{\rm min}=28^{+8}_{-11}$ $\text{km}\,\text{s}^{-1}$ and $v_{\rm max}=1189^{+450}_{-193}$ $\text{km}\,\text{s}^{-1}$, corresponding to a mean velocity of $308^{+83}_{-44}$ $\text{km}\,\text{s}^{-1}$.
As shown in Figure \ref{image_6}, the posterior distribution is broad, with a mode at 34 $\text{km}\,\text{s}^{-1}$, a 68\% highest density interval spanning from 34 to 354 $\text{km}\,\text{s}^{-1}$, and a 90\% credibility upper limit of 748 $\text{km}\,\text{s}^{-1}$.

\section{Discussion and Conclusions}\label{sec:6}

We have presented a hierarchical Bayesian analysis to reconstruct the intrinsic 3D velocity distribution of pulsars, incorporating the observational constraint of spin–velocity alignment. Using a carefully selected sample of 18 pulsars with measured proper motions, distances, and spin geometries, our framework consistently propagates measurement uncertainties into the population-level inference. Model comparison shows that a Gamma distribution yields the highest evidence among the candidates considered (BF = 1.65 relative to a single Maxwellian), although this small Bayes factor indicates that the present data do not provide meaningful discriminatory power between these models. The corresponding posterior parameters are $k=2.53^{+1.13}_{-0.81}$ and $\theta=156^{+80}_{-51}$ $\text{km}\,\text{s}^{-1}$.

To assess the robustness of our results, we performed sensitivity analyses addressing distance uncertainties and the treatment of Galactic potential corrections. Motivated by potential systematic biases in electron density models \citep{2019ApJ...875..100D}, we increased the assumed 20\% uncertainty for DM-based distances to 50\% and 100\%. Under these expanded uncertainty assumptions, the Gamma distribution continues to yield the highest evidence among the models considered, with Bayes factors of 4.61 and 3.68 relative to the Maxwellian distribution, respectively.
Although the Bayes factors increase slightly under broader distance uncertainties, the qualitative model ranking remains unchanged.
We also tested the impact of dynamical corrections by omitting backtracking for the seven sources aged 10–25 Myr. Even when using the observed velocities without Galactic potential corrections, the Gamma model again yields the highest evidence, with Bayes factors of 5.48 and 2.34 for the alignment-constrained and isotropic cases, respectively.
Overall, these tests indicate that the relative ordering of models is qualitatively stable with respect to distance uncertainties and Galactic potential modeling, although the Bayes factors remain modest and do not provide decisive discrimination between candidate distributions.

Crucially, the incorporation of the spin-velocity alignment constraint serves to mitigate the significant parameter degeneracies inherent in velocity reconstruction based solely on 2D data. Imposing the spin–velocity alignment constraint systematically lowers the reconstructed 3D velocities of individual pulsars relative to isotropic assumptions.
This suggests that projection effects under isotropic assumptions can bias individual velocity estimates high, even though the inferred population-level characteristic scale remains broadly consistent within uncertainties ($237^{+67}_{-84}$ $\text{km}\,\text{s}^{-1}$ vs $265^{+35}_{-42}$ $\text{km}\,\text{s}^{-1}$).

Finally, we address the apparent divergence between the Gamma distribution identified in our alignment-constrained sample and the Log-Normal distribution supported by the full sample of 465 pulsars (Appendix~\ref{sec:append}). We interpret these findings as complementary rather than contradictory, reflecting the dynamical evolution of the pulsar population. The Gamma-like structure observed in our young, geometry-constrained sample—also broadly consistent with the sub-population of isolated young pulsars in the full ATNF catalog—may more closely reflect the birth velocity distribution. In contrast, the Log-Normal profile observed in the broader population likely represents a dynamically evolved state, in which initial features have been smoothed by Galactic potential interactions and the admixture of recycled pulsars. Thus, robustly constraining natal kick physics requires disentangling intrinsic birth velocities from the effects of subsequent Galactic and stellar evolution.

This study is primarily limited by the current scarcity of pulsars with high-precision distance and full spin orientation measurements. Future high-precision parallax measurements from VLBI and next-generation radio astrometry, coupled with dedicated, high-cadence polarization observations, will dramatically expand the sample of pulsars with well-constrained spin-velocity geometry. Applying this hierarchical Bayesian framework to such an expanded dataset will provide the definitive statistical power needed to distinguish between competing natal kick theories.

\section{Acknowledgments}

This work is supported by the National Key Research and Development Program of China (No.~2023YFC2206704), the Fundamental Research Funds for the Central Universities, and the Supplemental Funds for Major Scientific Research Projects of Beijing Normal University (Zhuhai) under Project ZHPT2025001.
We thank the referee for very useful comments.
Xiaojin Liu is supported by the National Natural Science Foundation of China (No.~12503046).
Zhi-Qiang You is supported by the National Natural Science Foundation of China (Grant No.~12305059); The Startup Research Fund of Henan Academy of Sciences (No.~241841224); The Scientific and Technological Research Project of Henan Academy of Science (No.~20252345003); Joint Fund of Henan Province Science and Technology R\&D Program (No.~235200810111); Henan Province High-Level Talent Internationalization Cultivation (No.~20252645001).
Jumei Yao was supported by the National Science Foundation of Xinjiang Uygur Autonomous Region (2022D01D85), the Major Science and Technology Program of Xinjiang Uygur Autonomous Region (2022A03013-2), the Tianchi Talent Project, the CAS Project for Young Scientists in Basic Research (YSBR-063), the Tianshan Talents Program (2023TSYCTD0013), and the Chinese Academy of Sciences (CAS) ``Light of West China" Program (No.~xbzg-zdsys-202410 and No.~2022-XBQNXZ-015).

\section*{Data Availability}
The data underlying this article, including the analysis scripts and the curated pulsar datasets, are available in a GitHub repository at \url{https://github.com/Astro-ZL/Pulsar_Alignment_3D}.
\appendix
\twocolumngrid
\setcounter{figure}{0}
\setcounter{table}{0}
\renewcommand{\thefigure}{\thesection\arabic{figure}}
\renewcommand{\thetable}{\thesection\arabic{table}}

\section{Supplementary Model Comparison: The Velocity Distribution of the Full ATNF Pulsar Sample}
\label{sec:append}

We conduct a complementary hierarchical Bayesian analysis using transverse velocity data from the ATNF pulsar catalog to compare our geometry-constrained results with those derived from the larger population under the conventional isotropic assumption.

We begin with the full ATNF catalog and exclude pulsars in globular clusters and sources with transverse velocities exceeding $10^{4}\,\mathrm{km\,s^{-1}}$
. A 20\% Gaussian uncertainty is assumed for each pulsar distance. Following standard practice, we assume an isotropic distribution of velocity directions to convert 2D transverse velocities into 3D speed estimates.

\begin{figure}[h]
    \centering
         \includegraphics[scale=0.55]{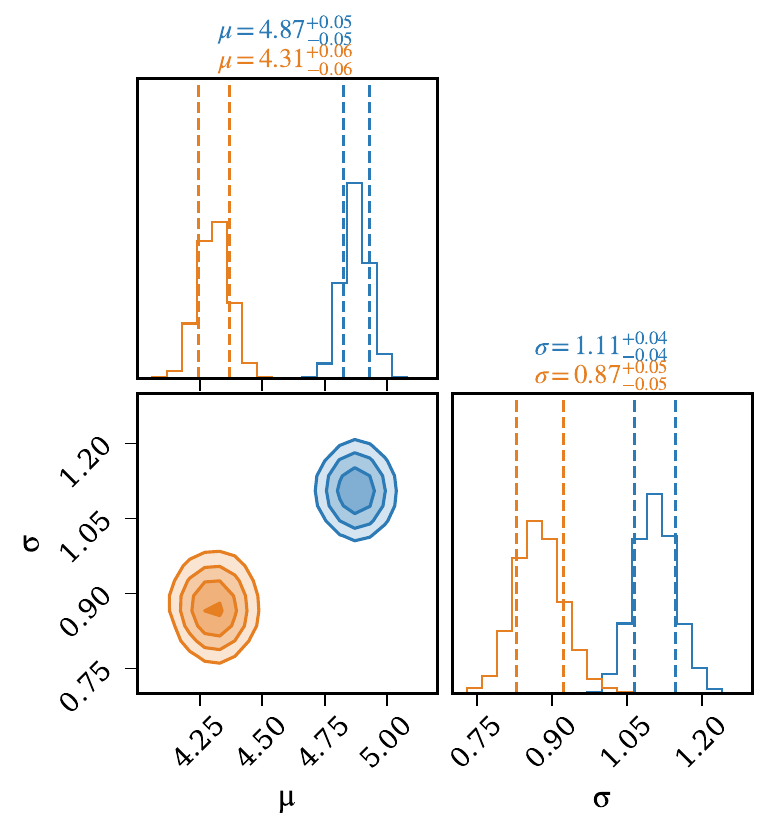}
         \caption{Posterior distributions of the log-normal model for the velocity distributions derived for the full sample (blue) and recycled pulsars (orange), with the dashed vertical lines marking the 1$\sigma$ confidence intervals.}
         \label{image_7}
\end{figure}

\begin{figure}[h]
    \centering
         \includegraphics[scale=0.55]{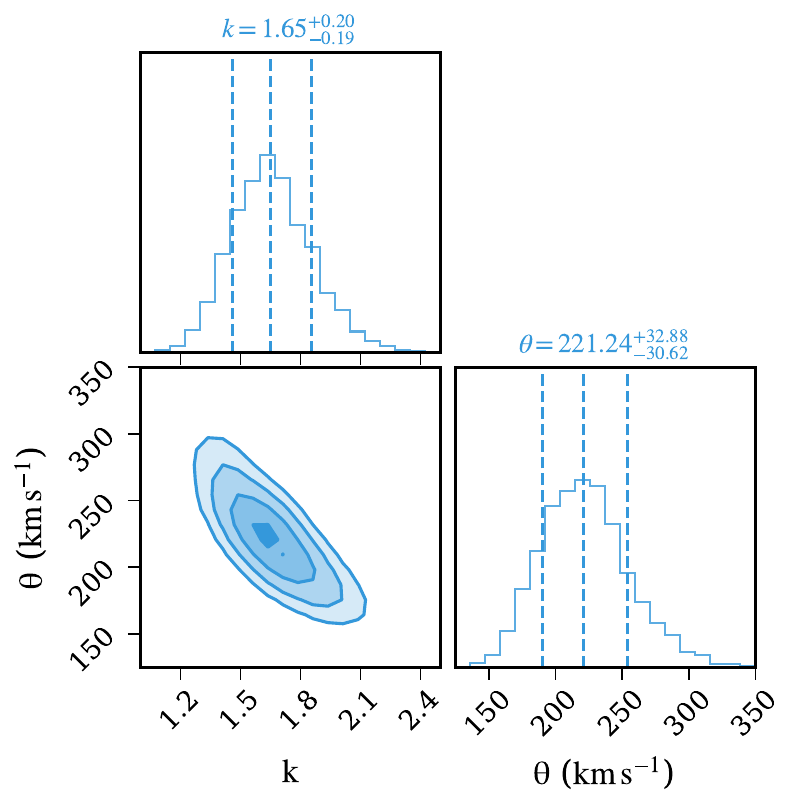}
         \caption{Posterior distributions of the gamma model for the velocity distribution of isolated pulsars, with the dashed vertical lines marking the 1$\sigma$ confidence intervals.}
         \label{image_8}
\end{figure}

The resulting full samples consists of 465 pulsars, and we further consider two subsets: 154 isolated young pulsars ($\tau_{\mathrm{c}}<10\,\mathrm{Myr}$), and 226 recycled pulsars ($P<0.1$ s and $\dot{P}<10^{-17}$).
We apply the same hierarchical Bayesian framework described in Section \ref{sec:4} to these datasets, evaluating the same nine parametric velocity distribution models.
Figures \ref{image_7} and \ref{image_8} display the posterior distributions for key models across populations.

The prior choices, posterior estimates, and log Bayes factors (lnBF) for all models relative to the single Maxwellian model are summarized in Table \ref{tab4}. Several key patterns emerge.
For the full sample, the lognormal distribution is decisively favored (lnBF = 398.1), with all other models strongly disfavored. This result aligns with recent findings by \cite{2025ApJ...989L...8D}.
For recycled pulsars, again, the lognormal model performs best (lnBF = 67.3), though the evidence over other models (particularly GA, 2G, and SST) is weaker than for the full sample.
For isolated young pulsars, the Gamma (GA) distribution shows the strongest preference (lnBF = 41.7), but the bimodal Gaussian (2G) model remains competitive (lnBF = 41.1). The difference is marginal, indicating that for young pulsars, both single-component skewed and two-component symmetric distributions are plausible.

Our velocity scales are systematically lower than those reported in recent isotropic analyses.
For isolated young pulsars, we find a mean 3D velocity of $366^{+107}_{-38}$ $\text{km}\,\text{s}^{-1}$ compared to $\sim 600$ $\text{km}\,\text{s}^{-1}$ in \citet{disberg2025kinematically}. This discrepancy likely arises from differences in distance uncertainties and sample selection.
The full-sample lognormal parameters ($\mu=4.87\pm 0.05$ and $\sigma=1.11 \pm 0.04$) correspond to a mean velocity of $242^{+54}_{-32}$ $\text{km}\,\text{s}^{-1}$, lower than the $\sim 345\,\mathrm{km\,s^{-1}}$ found in \cite{2025ApJ...989L...8D}.
The absence of strong bimodality in the full sample suggests that any multi-component velocity structure, if present in young pulsars, is diluted when older, recycled pulsars are included.
\twocolumngrid
\onecolumngrid
\begin{table*}[ht!]
\caption{Prior, 1$\sigma$ posterior intervals and lnBF of each model relative to the M model for 9 velocity-distribution models using the full sample, isolated-pulsars and recycled-pulsars samples.}
\label{tab4}
\small
\centering
\setlength{\tabcolsep}{2pt} 
\renewcommand{\arraystretch}{1.1} 
\begin{tabular}{ccccccccccc}
\hline\hline
\multirow{2}{*}{Model} &
\multirow{2}{*}{Parameters} &
\multicolumn{3}{c}{All pulsars} &
\multicolumn{3}{c}{Recycled pulsars} &
\multicolumn{3}{c}{Isolated pulsars} \\
\cline{3-11}
 & & Prior &Posterior&lnBF &  Prior&Posterior&lnBF &  Prior& Posterior&lnBF\\
\hline
\multirow{2}{*}{LN}
& $\mu$ & \text{U}[4,6] & 4.87$^{+0.05}_{-0.05}$ &\multirow{2}{*}{398.1}
&\text{U}[4,5]&4.31$^{+0.06}_{-0.06}$&\multirow{2}{*}{67.3}
&\text{U}[5,6]&5.58$^{+0.07}_{-0.08}$ &\multirow{2}{*}{36.7}\\
& $\sigma$ & \text{U}[0.5,1.5] & 1.11$^{+0.04}_{-0.04}$ &\multirow{2}{*}{}
& \text{U}[0.5,1.5]&0.87$^{+0.05}_{-0.05}$&\multirow{2}{*}{}
&\text{U}[0.5,1.5]&0.89$^{+0.07}_{-0.07}$ &\multirow{2}{*}{}\\
\hline
\multirow{3}{*}{SST}
&$\mu$& \text{U}[150,1000]&232$^{+14}_{-13}$&\multirow{4}{*}{390.2}
&\text{U}[50,150]&104$^{+7}_{-6}$&\multirow{4}{*}{65.3}
&\text{U}[1,500]&361$^{+26}_{-24}$&\multirow{4}{*}{37.6}\\
&$\sigma$& \text{U}[100,2500]&738$^{+618}_{-279}$&\multirow{4}{*}{}
&\text{U}[50,150]&88$^{+17}_{-10}$&\multirow{4}{*}{}
&\text{U}[1,500]&270$^{+22}_{-19}$&\multirow{4}{*}{}\\
&$\nu$ & \text{logU}[0.01,300]&48$^{+120}_{-34}$&\multirow{4}{*}{}
&\text{logU}[0.01,300]&35$^{+112}_{-27}$&\multirow{4}{*}{}
&\text{logU}[0.01,300]&32$^{+118}_{-24}$&\multirow{4}{*}{}\\
&$\tau$ & \text{U}[2.01,10]&2.10$^{+0.15}_{-0.07}$&\multirow{4}{*}{}
&\text{U}[2.01,40]&5.40$^{+5.71}_{-1.82}$&\multirow{4}{*}{}
&\text{U}[2.01,400]&167$^{+156}_{-135}$&\multirow{4}{*}{}\\
\hline
\multirow{5}{*}{2G}
&$\mu_{1}$ &\text{U}[10,150]& 27$^{+20}_{-13}$&\multirow{5}{*}{386.7}
&\text{U}[1,100]&51$^{+14}_{-27}$&\multirow{5}{*}{63.4}
&\text{U}[1,300]&139$^{+88}_{-96}$&\multirow{5}{*}{41.1}\\
&$\sigma_{1}$&\text{U}[100,300]& 177$^{+18}_{-20}$&\multirow{5}{*}{}
&\text{U}[1,400]&75$^{+79}_{-22}$&\multirow{5}{*}{}
&\text{logU}[1,1000]&432$^{+175}_{-217}$&\multirow{5}{*}{}\\
&$\mu_{2}$&\text{U}[10,1000]& 306$^{+321}_{-197}$&\multirow{5}{*}{}
&\text{U}[1,500]&147$^{+140}_{-84}$&\multirow{5}{*}{}
&\text{logU}[1,1000]&247$^{+177}_{-45}$&\multirow{5}{*}{}\\
&$\sigma_{2}$&\text{U}[400,1800]& 816$^{+207}_{-160}$&\multirow{5}{*}{}
&\text{U}[1,500]&167$^{+132}_{-95}$&\multirow{5}{*}{}
&\text{logU}[1,1000]&226$^{+372}_{-213}$&\multirow{5}{*}{}\\
&$\alpha$ &\text{U}[0.01,0.99]& 0.87$^{+0.04}_{-0.05}$&\multirow{5}{*}{}
&\text{U}[0.01,0.99]&0.83$^{+0.11}_{-0.37}$&\multirow{5}{*}{}
&\text{U}[0.01,0.99]&0.73$^{+0.15}_{-0.45}$&\multirow{5}{*}{}\\
\hline
\multirow{3}{*}{E}
& $v_{\text{min}}$ & \text{U}[1,10] &2.49$^{+1.32}_{-1.05}$&\multirow{3}{*}{377.5}
&\text{U}[1,10]&2.57$^{+1.40}_{-1.08}$&\multirow{3}{*}{54.4}
&\text{U}[1,10]&4.20$^{+2.50}_{-2.16}$&\multirow{3}{*}{36.1}\\
& $v_{\text{max}}$ & \text{logU}[10,10000]&4644$^{+3204}_{-2026}$&\multirow{3}{*}{}
&\text{logU}[10,10000]&2674$^{+3954}_{-1621}$&\multirow{3}{*}{}
&\text{logU}[10,10000]&4740$^{+3204}_{-2000}$&\multirow{3}{*}{}\\
& $\lambda$  & \text{U}[0.004,0.005] & 0.0045$^{+0.0002}_{-0.0002}$&\multirow{3}{*}{}
&\text{logU}[0.001,0.1]&0.0093$^{+0.0007}_{-0.0006}$&\multirow{3}{*}{}
&\text{U}[0.002,0.004]&0.0027$^{+0.0002}_{-0.0002}$&\multirow{3}{*}{}\\
\hline
\multirow{2}{*}{GA} 
& k & \text{logU}[0.1,5] & 1.06$^{+0.07}_{-0.07}$&\multirow{2}{*}{374.4}
&\text{U}[1,5]&1.65$^{+0.18}_{-0.16}$&\multirow{2}{*}{66.7}
&\text{U}[1,5]&1.65$^{+0.20}_{-0.19}$&\multirow{2}{*}{41.7}\\
& $\theta$ & \text{U}[10,300] & 213$^{+19}_{-18}$&\multirow{2}{*}{}
&\text{U}[10,100]&63$^{+8}_{-7}$&\multirow{2}{*}{}
&\text{U}[10,400]& 221$^{+33}_{-31}$&\multirow{2}{*}{}\\
\hline
\multirow{2}{*}{LU} 
& $v_{\text{min}}$ & $\text{U}[1,50]$ &16.27$^{+1.39}_{-1.32}$&\multirow{2}{*}{340.6}
&\text{U}[1,50]&14.24$^{+1.79}_{-1.68}$&\multirow{2}{*}{41.9}
&\text{U}[1,100]&51$^{+9}_{-7}$&\multirow{2}{*}{12.1}\\ 
& $v_{\text{max}}$ & $\text{U}[50,10000]$ & 1500$^{+37}_{-21}$&\multirow{2}{*}{}
&\text{U}[50,10000]&464$^{+14}_{-9}$&\multirow{2}{*}{}
&\text{U}[100,10000]&1575$^{+61}_{-27}$&\multirow{2}{*}{}\\
\hline
\multirow{2}{*}{G}
& $\mu$ & \text{U}[10,500] & 14.63$^{+7.18}_{-3.38}$&\multirow{2}{*}{326.0}
&\text{U}[1,100]&17.28$^{+18.63}_{-11.97}$&\multirow{2}{*}{60.5}
&\text{U}[1,300]&99$^{+81}_{-65}$&\multirow{2}{*}{40.9}\\
& $\sigma$&\text{U}[10,500] &295$^{+13}_{-12}$&\multirow{2}{*}{}
&\text{U}[10,200]&121$^{+9}_{-11}$&\multirow{2}{*}{}
&\text{U}[100,500]&406$^{+45}_{-47}$&\multirow{2}{*}{}\\
\hline
\multirow{3}{*}{2M}
&$\sigma_{1}$& \text{U}[1,100]& 56$^{+4}_{-3}$&\multirow{3}{*}{320.8}
&\text{U}[1,60]&35$^{+5}_{-6}$&\multirow{3}{*}{53.7}
&\text{U}[1,200]&132$^{+18}_{-18}$&\multirow{3}{*}{29.7}\\
&$\sigma_{2}$&\text{U}[100,400]& 301$^{+20}_{-17}$&\multirow{3}{*}{}
&\text{U}[60,200]&115$^{+19}_{-15}$&\multirow{3}{*}{}
&\text{U}[200,600]&391$^{+58}_{-45}$&\multirow{3}{*}{}\\
&$\alpha$& \text{U}[0.01,0.99]& 0.64$^{+0.03}_{-0.03}$&\multirow{3}{*}{}
&\text{U}[0.01,0.99]&0.62$^{+0.10}_{-0.12}$&\multirow{3}{*}{}
&\text{U}[0.01,0.99]&0.62$^{+0.11}_{-0.11}$&\multirow{3}{*}{}\\
\hline
\multirow{1}{*}{M}
&$\sigma$&\text{U}[10,200]&154$^{+4}_{-4}$&\multirow{1}{*}{0}
&\text{U}[10,100]&66$^{+3}_{-3}$&\multirow{1}{*}{0}
&\text{U}[10,300]&235$^{+11}_{-10}$&\multirow{1}{*}{0}\\
\hline
\end{tabular}
\end{table*}


\twocolumngrid 
\bibliography{kick}{}
\bibliographystyle{aasjournal}

\end{document}